\documentclass[runningheads]{llncs}
\usepackage{graphicx}
\usepackage{xcolor}
\usepackage{hyperref}
\usepackage[english]{babel}

\begin{document}
\title{Verification and Validation of Computer Models for Diagnosing Breast Cancer Based on Machine Learning for Medical Data Analysis \thanks{AL and VL are grateful to Russian Science Foundation (grant RFBR No. 19-01-00358) for the financial support of the development of mathematical models for early diagnosis of breast cancer. MP is grateful to RFBR and the government of Volgograd region according to the research project No. 19-47-343008 for the financial support.  AK acknowledges Ministry of Science and Higher Education of the Russian Federation (government task, project No.\,2.852.2017/4.6) for the financial support of the development of the software and numerical simulations.}}
%
%
\author{Vladislav Levshinskii \and
Maxim Polyakov\and
Alexander Losev
\and Alexander Khoperskov
}
%
\authorrunning{V. Levshinskii et al.}
%
\institute{Volgograd State University, Volgograd, Russia \\
\email{\{v.levshinskii, m.v.polyakov, alexander.losev, khoperskov\}@volsu.ru}
 }
%
\maketitle              
\begin{abstract}
 The method of microwave radiometry is one of the areas of medical diagnosis of breast cancer. It is based on analysis of the spatial distribution of internal and surface tissue temperatures, which are measured in the microwave (RTM) and infrared (IR) ranges. Complex mathematical and computer models describing complex physical and biological processes within biotissue increase the efficiency of this method. Physical and biological processes are related to temperature dynamics and microwave electromagnetic radiation. Verification and validation of the numerical model is a key challenge to ensure consistency with medical big data. These data are obtained by medical measurements of patients. We present an original approach to verification and validation of simulation models of physical processes in biological tissues. Our approach is based on deep analysis of medical data and we use machine learning algorithms. We have achieved impressive success for the model of dynamics of thermal processes in a breast with cancer foci. This method allows us to carry out a significant refinement of almost all parameters of the mathematical model in order to achieve the maximum possible adequacy.\footnote{\color{red}Levshinskii V., Polyakov M., Alexander Losev A., Khoperskov A.V. Verification and Validation of Computer Models for Diagnosing Breast Cancer Based on Machine Learning for Medical Data Analysis // Communications in Computer and Information Science, 2019, v. 1084, pp. 447-460 \ \ \href{https://doi.org/10.1007/978-3-030-29750-3\_35}{https://doi.org/10.1007/978-3-030-29750-3\_35} }

\end{abstract}
\section{Introduction}
Today a very relevant problem is the creation of diagnostic technologies based on the integration of modern engineering developments, the most recent medical knowledge and the latest developments in the fields of mathematical modeling and machine learning. Foremost it is connected with the demand for the creation of a brand new medical equipment, as well as the creation of functional diagnostics methods based on dynamic measurements, description and interpretation of the parameters of physical fields and radiation of human body. At the same time, we must note one of the most difficult and relevant problem in medicine. It is a problem of early differential diagnosis of breast diseases. According to statistics, breast cancer is the most commonly occurring cancer in women. In 2012 more than 1.7 million cases of female breast cancer were diagnosed worldwide, which is about 12\,\% of all types of cancer. But the forecast for 2025 is more shocking, because 19.3 million new cases of cancer will be registered \cite{WHO2013WorldCancerStatistics,Bray2013GlobalEstimatesCancer}.

The problem of diagnostics is far away from any acceptable solution despite the considerable achievements over the past decades in understanding the clinical picture of the disease, the presence of major changes in treatment approaches, and notable development of functional diagnostics methods. The problem statement of survivability enhancement through early detection and subsequent appropriate treatment seemed quite obvious. However, the actual results, at the moment, is very far from the expected. Traditional methods that are currently in use do not effectively detect a tumor at an early stage. There are two complementary ways to solve this problem. The first one based on the usage of a complex of methods in early diagnostics. The second one is the development and improvement of new diagnostic methods. At the same time, the current situation in the world is such that in most cases the modern medical equipment application after solving some problems gives rise to others. In most cases difficulties in diagnostics arise not because of a lack of information, but because of underperformance of its processing methods. To some extent, the creation of systems for medical data mining and interpretation provides a solution to the said problems. Using methods and algorithms of machine learning, such systems should help specialists in the diagnostics, prediction of disease development, etc.

The application of information technologies in the diagnostics of diseases, as well as breast cancer, is not a new idea. For example, administrative databases can be used for early detection of breast cancer \cite{Abraha2018administrativeDatabases}. In the paper \cite{Igali2018ThermalModelingBreast}, the authors proposed a comprehensive method that includes FEM modelling based on heat transfer principles and requires 3D scanning and IR imaging for disease diagnostics. Naturally, various algorithms and methods of machine learning had been showing a sufficiently high efficiency in medicine \cite{Mohanty2013DataMiningTechnique,Yassin2018MachineLearning,Horsch2011ImageDatabases,LosevLevshinskiy2017Vestnik,Zenovich2018AlgorithmsDiagnosis}. Many studies are devoted to the analysis of the effectiveness of various decision support systems \cite{Beeler2014DSS,Miller2016DSS,Manar2017DSS,Wasylewicz2019DSS,Walsh2019DSSOncology}.

Microwave radiothermometry is one of the most promising methods for increasing the effectiveness of breast screening and early differential diagnosis. It is known that human tissues, like any heated body, emit electromagnetic oscillations in a wide frequency range. In addition, the radiation spectrum and the intensity of electromagnetic radiation of tissues in the microwave range is determined not only by the temperature distribution, but also by absorption and reradiation in physically inhomogeneous tissues. Thus, by measuring this radiation, one can obtain an additional and extremely important information about the state of internal and skin tissues. This is the main idea of the microwave radiothermometry method that based on measurements of tissue radiation in the microwave (RTM) and infrared (IR) wavelength ranges. It is known that changes in tissue temperature usually precede structural changes that are found with traditional diagnostic methods. In particular, temperature anomalies can be caused by enhanced metabolism of cancer cells. Early diagnosis of breast cancer is based on this characteristic.

For the first time ever the possibility of using microwave radiothermometry for the diagnosis of breast cancer was shown in \cite{Bar1}. To some extent, it can be considered that the theoretical basis of microwave radiothermometry in mammology is founded on the studies of the French scientist M. Gautherie \cite{Gau1}. Based on the clinical data of more than 85,000 patients, he convincingly showed that the heat radiation of a tumor is directly proportional to the rate of its growth. Thus, microwave radiothetmometry has a unique ability to detect fast-growing tumors in the first place.

In general, despite the obvious advantages of microwave radiothermometry, for a long time it has not received proper application in medical practice and worthy attention in scientific research. As mentioned above, this is due to the complexity of creating measuring equipment and the quality of measurements. Interest in this technique has significantly grown after the creation of effective microwave radiometers \cite{LosevLevshinskiy2017Vestnik,Sedankin2018Antenna,Avila-Castro2017}. Note that microwave radiothermometry is applicable not only for the diagnosis of breast cancer, but also for the diagnosis of other diseases \cite{Sedankin2018MathematicalSimulation}.

The main task of our project is the development and application of the method of iterative verification and validation of thermometric data obtained with simulation models. The objective is to find a correct range of the mathematical model parameters based on machine learning technologies and statistical data analysis. The proposed method will allow the development of adequate computer physical and mathematical models for studying the spatial and temporal dynamics of temperature and radiation fields in biological tissues of mammary gland.

\section{Thermal model of breast and model RTM-measurements}

\subsection{Model of breast}
The breast has a complex, multicomponent, heterogeneous structure. We distinguish between muscle tissue, adipose tissue, blood flow, breast lobes, areola, nipple, excretory ducts, connective tissue, skin. Effect for cancer is a separate problem because the location and physical characteristics of the tumor are unknown.

\begin{figure} [!htpb]
\centering
	\includegraphics[width=0.5\textwidth]{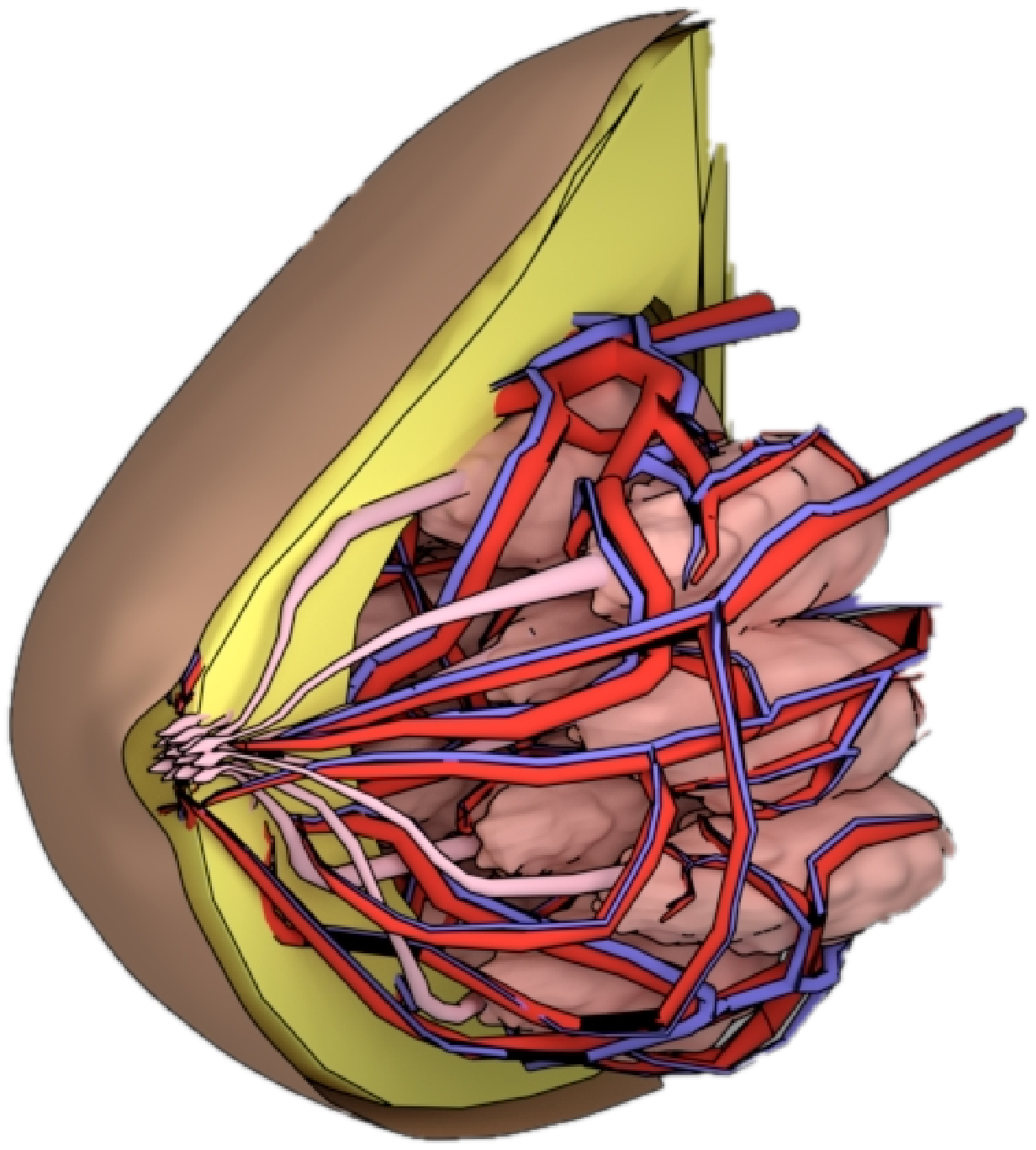}
\includegraphics[width=0.3\textwidth]{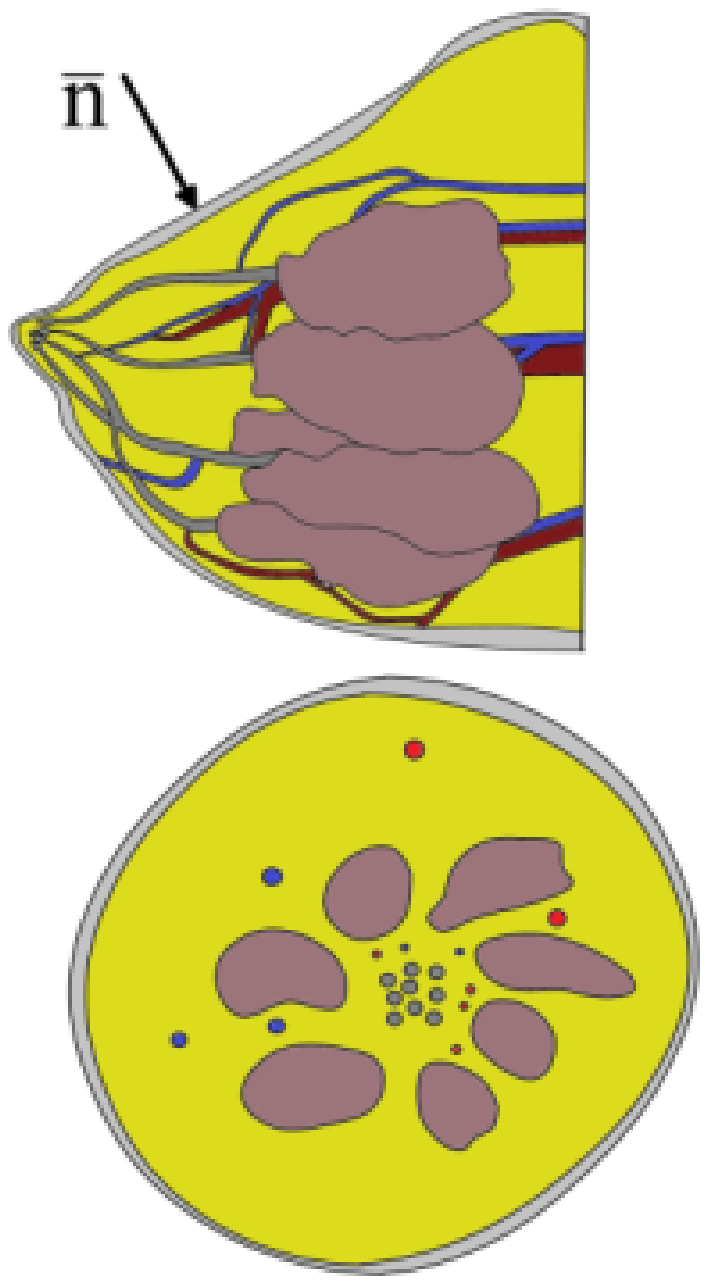}
	\caption{3D model of the breast. We show the internal small-scale structure of biotissue. Breast model slices are shown in different planes} \label{figBreast01}
\end{figure}

 These components are characterized by complex geometry, strong heterogeneity of their physical and chemical characteristics $\vec{G}(\vec{r})$ and individual variations of physical parameters $\delta{\vec{G}}$. We determine the state of biotissue using a vector of physical characteristics.
 \begin{equation}\label{tuple}
 \vec{G}(\vec{r}) \rightleftharpoons \langle \varrho, c_p, \lambda, \varepsilon_c, Q_{met}, Q_{bl}, Q_{car}, Q_{rad}, h_{air}, \sigma, \vec{n}_s, ... \rangle ,
  \end{equation}
  where $\varrho$ is a density, $c_p$ is a specific heat, $\lambda$ is a thermal conductivity coefficient, $\varepsilon_c$ is a  complex permittivity, $Q_{met}$, $Q_{bl}$, $Q_{car}$ are the rate of heat release from metabolic processes, blood flows and cancers, respectively, $Q_{rad}$ is a radiative cooling, $h_{air}$ is a  convective heat transfer coefficient, $\sigma$ is a electrical conductivity, $\vec{n}_s$ is a vector that determines the characteristic shape of the breast. Each characteristic depends on the coordinates inside the biotissue.

  \begin{figure} [!htpb]
\centering
	\includegraphics[width=0.95\textwidth]{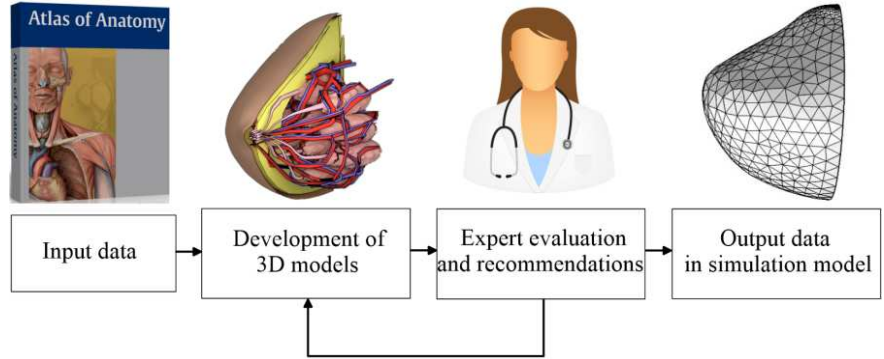}
	\caption{Scheme of procedure for building a 3D model of the breast} \label{figBreast02}
\end{figure}

Individual variations in the properties of biotissue components can be significant, amounting normally to tens of percent for some parameters $G_j$.
 We use a more realistic geometric structure of tissues with heterogeneous characteristics instead of the traditionally used models with homogeneous parameters in a multilayer approximation (usually limited to four types of tissues --- skin, muscles, mammary glands, tumor) \cite{Sedankin2018MathematicalSimulation}. We also take into account the filamentous connective tissues, the share structure of the breast, areola, nipple, excretory ducts (Fig.\,\ref{figBreast01}).

 The procedure for constructing a 3D model of the internal structure of the breast is shown on fig.\,\ref{figBreast02}.

\subsection{Thermal model}

The process of heat transfer in biotissue is described by the equation \cite{Polyakov2017NumericalModeling}
\begin{equation}\label{eq-heat-dynamics}
 \varrho(\vec{r}) c_p(\vec{r})\, \frac{\partial T}{\partial t} = \vec{\nabla} \left( \lambda(\vec{r}) \vec{\nabla} T  \right) + \sum\limits_{j} Q_j(\vec{r})  \,,
\end{equation}
\begin{equation}\label{eq-heat-dynamics-boundary}
\vec{n}(\vec{r})\cdot \vec{\nabla T} = \frac{h_{air}}{\lambda(\vec{r})} \cdot \left( T - T_{air} \right)
\ \ \ {\textrm{for}}\ \ \vec{r}\in \vec{S}_0(\vec{r})=\vec{n}(\vec{r})\cdot S_0(\vec{r}) \,,
\end{equation}
where $T(\vec{r})$  is the temperature at the point $\vec{r}=\left\{ x,y,z \right\}$, $\vec{\nabla}$ is differential operator nabla, heat source $Q_j > 0$ for $j=\{ met,bl,car \}$ and $Q_{rad}<0$, $T_{air}$ is temperature of environment, $\vec{S}_0(\vec{r})$ is the boundary of biological tissue, $\vec{n}$ is unit normal vector (see Fig.\,\ref{figBreast01}) \cite{Avila-Castro2017,Polyakov2017NumericalModeling}. The equation specifies the boundary conditions between biological tissue and environment.

\begin{figure}
\centering
	\includegraphics[width=1.0\textwidth]{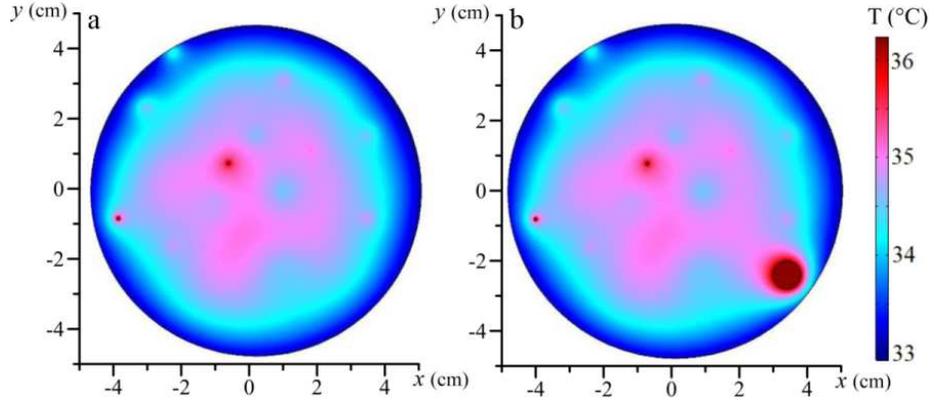}
	\caption{Distribution of the thermodynamic temperature $T$ at a depth of 3\,cm: without cancer (a), with cancer (see right lower part) (b)} \label{figBreast03}
\end{figure}

\subsection{Modeling of the radiation field and numerical model}

An important problem of the microwave radiometry method is the difference between the measured temperature (brightness temperature) $T_B$ and the thermodynamic temperature $T$ (see Fig.\,\ref{figBreast03}). Measurements are carried out using microwave antennas.

\begin{figure}
\centering
\includegraphics[width=1.0\textwidth]{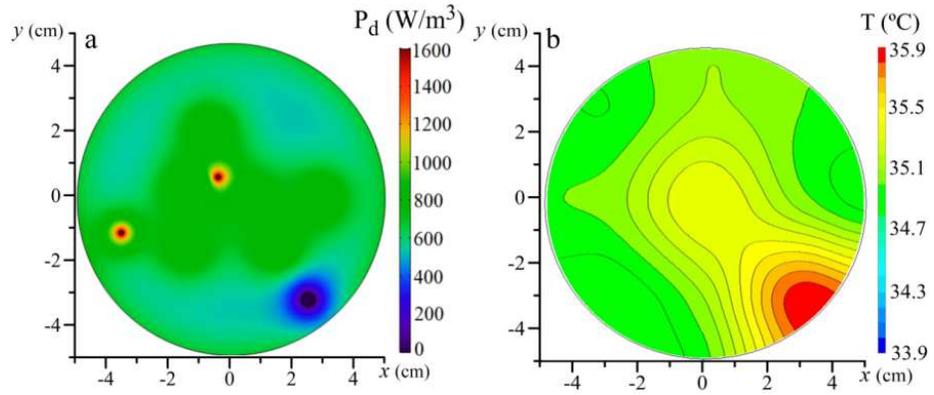}
	\caption{Distribution of $P_d$ (a). Brightness temperature distribution $T_B$ (b)} \label{figBreast04}
\end{figure}

Brightness temperature equals to
\begin{eqnarray}
 T_B(\vec{r}) = \int\limits_{\Delta{f}} \Biggl\{ \left| S_{11}(f) \right|^2 T_{REC} + \left[ 1 - \left| S_{11}(f) \right|^2 \right] \times
 \nonumber
 \\
 \times  \left( \int_{V_{0}} T(\vec{r}) \,\frac{P_d(\vec{r},f)}{\int_{V_0}P_d(\vec{r},f)\,dV}\,dV + T_{EMI} \right) \Biggr\}\, df \,,
 \label{eq-TBintegral}
\end{eqnarray}
where $\displaystyle P_d = \frac{1}{2}\,\sigma(\vec{r},f)\cdot \left| \vec{E}(\vec{r},f) \right|^2$ is electromagnetic field power density (see Fig.\,\ref{figBreast04}(a)), $\vec{E}$ is an electric field vector. Values $T_{EMI}$ and $T_{REC}$ characterize electromagnetic interference when measured with a radiometer \cite{Sedankin2018Antenna,Sedankin2018MathematicalSimulation}. The coefficient $S_{11}$ determines the interaction between the antenna and the biological tissue. Integration is carried out over the entire volume of biotissue ($V_0$).
Frequency range $\Delta{f}=1.3\div 1.5$\,GHz.

To construct a stationary electric field distribution, it is convenient to solve the time-dependent Maxwell equations and as the result to obtain the stationary-state:
\begin{equation}\label{eq-Makswell}
    \frac{\partial \vec{B}}{\partial t} + rot(\vec{E}) = 0 \,, \quad \frac{\partial \vec{D}}{\partial t} - rot(\vec{H}) = 0 \,,\quad \vec{B}=\mu\vec{H}\,,\quad \vec{D}=\varepsilon\vec{E} \,,
\end{equation}
where  $\vec{B}$ is magnetic induction, $\vec{E}$ is electric field strength, $\vec{D}$ is electric induction, $\vec{H}$ is magnetic field strength, $\varepsilon(\vec{r})$ is the dielectric constant, $\mu(\vec{r})$ is magnetic permeability.

Procedure of numerical solution for determination of internal temperature distribution $T_B(x,y,z)$ inside the tissue volume in the model (\ref{eq-heat-dynamics})--(\ref{eq-TBintegral}) described in \cite{Polyakov2017NumericalModeling}.
 The value $T_B(\vec{r},t)$ (see Fig.\,\ref{figBreast04}(b)) is a convolution of the thermodynamic temperature distribution $T(\vec{r},t)$.  It is obtained by solving the problem (\ref{eq-heat-dynamics})--(\ref{eq-heat-dynamics-boundary}), and electric field distribution $\vec{E}(\vec{r},t)$.

The development of new medical antennas and radiometers with improved performance is an important practical problem \cite{Sedankin2018MathematicalSimulation}.
The method of microwave radiometry requires high precision temperature measurement. It allows us to detect thermal anomalies non-invasively at a depth of several centimeters and localize the measurement point.

\begin{figure} [!htpb]
\centering
	\includegraphics[width=0.6\textwidth]{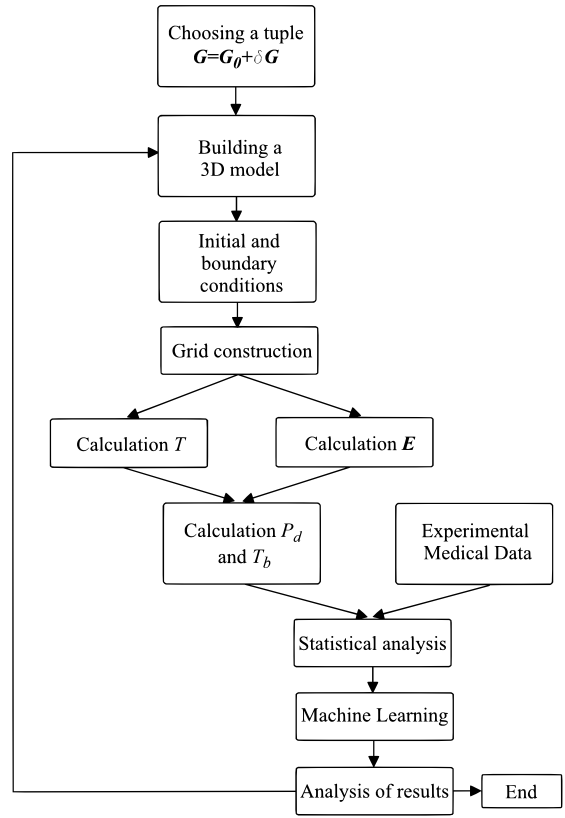}
	\caption{Algorithm for conducting numerical experiment and verification}
 \label{Algorithm}
\end{figure}

The emergence printed antennas for microwave radiothermometry (see \cite{Sedankin2018Antenna}) improve the efficiency of diagnosis based on an analysis of the internal temperature of the breast.
Switching to textile antennas, that are embedded in clothing and are able to conduct continuous monitoring can make revolutionary advances in breast oncology practice.

\section{Numerical Model Validation Problem}

We make sequences with different $\vec{G}$ tuples to perform the procedure for verification and validation of computer simulation results. Each set contains $k = 80$ solutions to problems (2) - (5). Each implementation of the $F_k$ model ($k = 1, ..., K$) is determined by a random variation $\delta\vec{G}$ except tumors. The vector of physical characteristics has a natural variation. At the end of each iteration we had performed an analysis of modelled thermometric data and corrected the physical components. The algorithm for iterative verification is shown in figure \ref{Algorithm}.

\subsection{Radiothermometric breast examination and dataset}

\begin{figure}[!htb]
	\centering
	\includegraphics[width=0.95\textwidth]{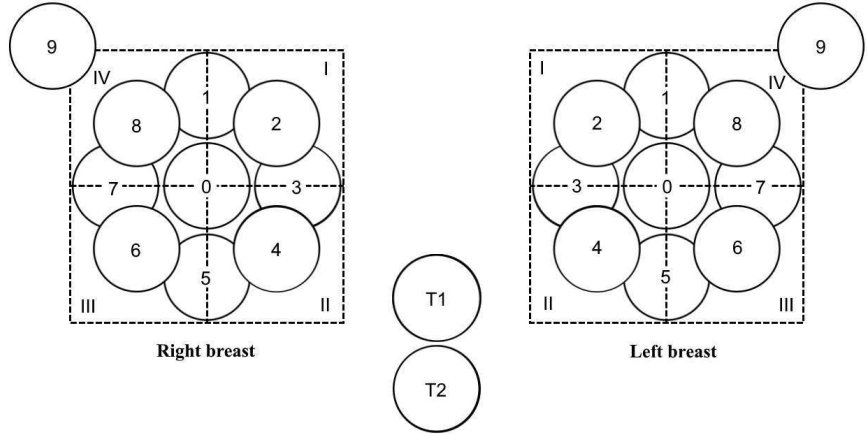}
	\caption{Radiothermometric examination chart} \label{fig:examScheme}
\end{figure}

The radiothermometric examination consists of the sequential measurement of surface and internal temperatures at certain points of the mammary glands and the subsequent recording of temperatures in the numerical form. The examination chart is shown in the figure \ref{fig:examScheme}: points $0, \dots, 8$ are located on the mammary gland, point $9$ is located in the axillary region, and points $T1$ and $T2$ are located between the mammary glands.

\begin{figure}[!htb]
	\centering
	\includegraphics[width=0.5\textwidth]{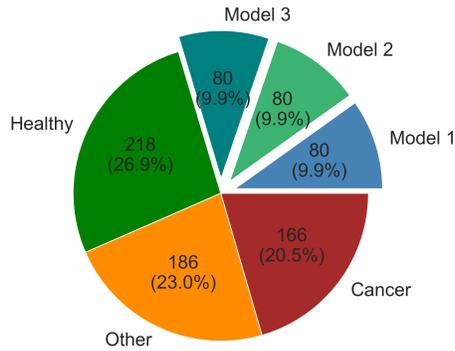}
	\caption{Per-class counts of breasts in the combined dataset} \label{fig:classesAll}
\end{figure}

We had formed a thermometric dataset based on the data from cancer centers. It contains the following data: temperature values, age, breast diameter, parity, etc. However, we need only consider temperature values at the points $0, \dots, 8$, because computer simulation results contain only points of the mammary gland. Each patient is assigned to a certain class from the following: healthy, other (various diseases of the mammary glands), cancer (malignant tumor). Next, the data set was combined with the simulation results. The number of mammary glands per class is shown in the figure \ref{fig:classesAll}. For clarity, the simulation results obtained after the first and second iterations of the verification algorithm are also shown.

We denote the combined dataset as
\begin{equation}\label{eq:dataset}
X =
\left[ \matrix{
t^{1}_{0} & t^{1}_{1} & \dots & t^{1}_{18} \cr
t^{2}_{0} & t^{2}_{1} & \dots & t^{2}_{18} \cr
\dots & \dots & \dots & \dots \cr
t^{810}_{0} & t^{810}_{1} & \dots & t^{810}_{18}
\cr} \right],
y =
\left[ \matrix{
y_{1} \cr
y_{2} \cr
\dots \cr
y_{810}
\cr} \right]
\end{equation}
where $t^i_0, \dots, t^i_{8}$ are internal temperatures at the points $0, \dots, 8$, $t^i_9, \dots, t^i_{18}$ are surface temperatures and $y_i \in \{\textrm{Model 1}, \textrm{Model 2}, \textrm{Model 3}, \textrm{Healthy}, \textrm{Other}, \textrm{Cancer}\}$ is a label of the $i$-th breast.

\subsection{Feature Space}

The original feature space contains the values of surface and internal temperatures. However, at the anomaly detection process the experts analyze not the temperature values, but their various relations. Based on this fact in the process of thermometric data mining, we proposed several hypotheses about the behaviour of temperature fields and significantly expanded the feature space. It allows to detect temperature anomalies with better performance and to build various classification algorithms \cite{LosevLevshinskiy2017Vestnik,Zenovich2018AlgorithmsDiagnosis,ZenovichGrebnevPrimachenko2017NN} that can provide a full explanation of the result and underlie the intelligent system for diagnostics of the mammary glands diseases.

Taking the above into account, the feature space of the combined dataset was extended by the following features:
\begin{enumerate}
	\item The values of functions of the form $T^{g}_i = t_{i} - t_{i + 9}, i = \overline{0,8}$ that are the so-called internal temperature gradients at the points $0, \dots, 8$. These functions represent temperature changes with respect to depth;
	\item Values of functions of the form $R^{mw}_i = t_{0} - t_{i}, i = \overline{1,8}$ and $R^{ir}_i = t_{9} - t_{i}, i = \overline{10,18}$ that are so-called radial gradients. These functions represent temperature changes relative to nipple temperature, which is the one of the most important features;
	\item The values of functions of the form $G^2 = T^{g}_0 - T^{g}_i, i = \overline{1,8}$ that describe changes of internal temperature gradients in radial direction. These functions are the difference analogs of the second derivatives of temperature functions \cite{LosevLevshinskiy2017Vestnik};
	\item More general forms of functions from other groups. For example, the maximum value of the internal gradients $F_1 = \max\limits_{i \in \overline{0, 8}}\left|T^{g}_i\right|$.
\end{enumerate}

\subsection{Verification and Validation}

We apply the following algorithm to verify the simulation results:
\begin{enumerate}
	\item Build a classifier using the real data and test it on the model data;
	\item Build a classifier using the model data and test it on the real data;
	\item Analyze the results and the values of features for the wrong predictions;
	\item Change the model coefficients, simulate new data and return to step 1.
\end{enumerate}

Consider the breast cancer binary classification problem. Let patients with breast cancer be positive examples and patients without breast diseases (or computer simulation results) be negative examples. Note that we do not consider a group of patients called "Other" and left it in the figures for clarity. This is due to the complexity and diversity of the diseased breast structure.

As the model evaluation criterion we use the following
\begin{equation}\label{eq-Eval}
M = \sqrt{ Spec_1 \cdot Spec_2} \,,
\end{equation}
where $\displaystyle Spec_1$ is the specificity of a classifier built at step 1 and tested on the model data, $\displaystyle Spec_2$ is the specificity of a classifier built at step 2 and tested on the real data, $\displaystyle Spec = \frac{TN}{TN + FP}$ is the specificity, $TN$ is the number of true negatives and $FP$ is the number of false positives.

As the classifier evaluation criterion we use the following \begin{equation}\label{eq-Sens-Spec}
G = \sqrt{Sens \cdot Spec} \,,
\end{equation}
where $\displaystyle Sens = \frac{TP}{TP + FN}$ is the sensitivity (recall), $TP$ is the number of true positives and $FN$ is the number of false negatives.

As a classifier we use the weighted simple rule voting algorithm, that is a modified version of an algorithm proposed in \cite{LosevLevshinskiy2017Vestnik}. Its performance measure reported by 5-fold nested cross-validation is 0.83, and specificity is close to 1 after refitting. A key feature of this classifier is the fully interpretable output that allows to immediately detect differences in patterns.

\subsection{Results and Analysis}

After the first iteration of the verification algorithm, we measured $M = 0.67$, $\displaystyle Spec_1 = 0.75$ and $\displaystyle Spec_2 = 0.61$. The analysis of the results has shown that the surface temperatures in modelled data is characterized by the lower values. In addition, the modelled data has a higher variance.

\begin{figure}
	\includegraphics[width=0.99\textwidth]{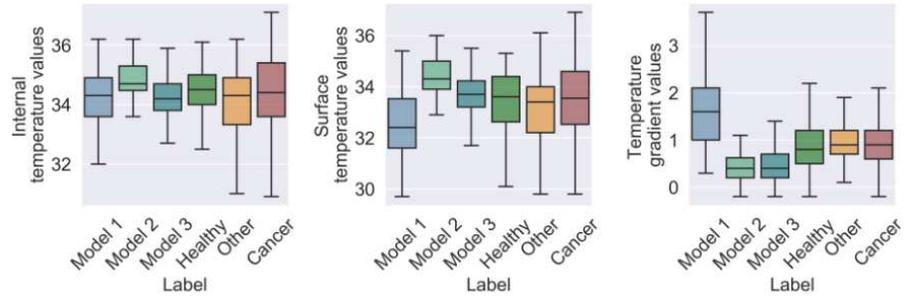}
	\caption{Boxplots of temperature values at the point 0 (outliers aren't shown)} \label{fig:boxPlot}
\end{figure}

\begin{figure}
	\includegraphics[width=0.99\textwidth]{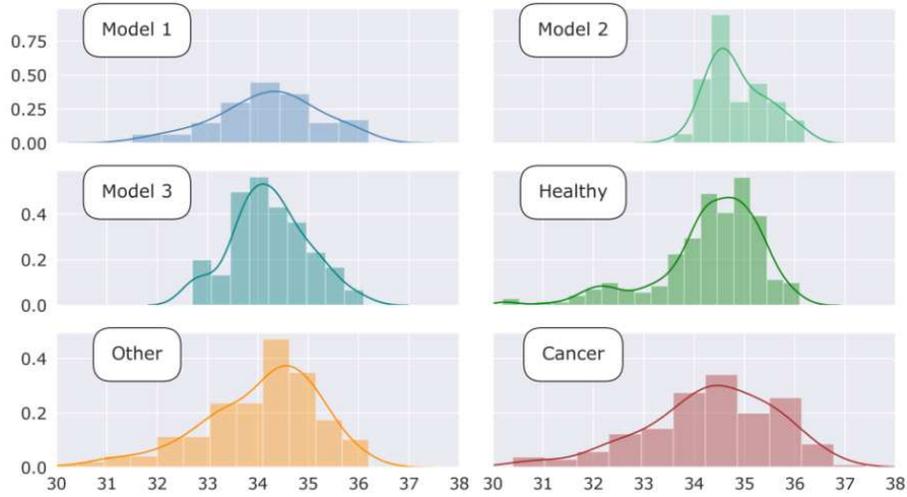}
	\caption{Per-class distributions of the internal temperature values at the point 0} \label{fig:distPlot}
\end{figure}

The problems mentioned above were taken into account at subsequent iterations of the algortihm. At the steps 2 and 3 we measured nearly identical $M = 0.72$. The second modelling results are characterized by higher surface and internal temperature values, as well as low variance. At the third iteration, these gaps were partially eliminated. Typical situations are shown in the figures \ref{fig:boxPlot} and \ref{fig:distPlot}.

\section{Conclusion}
Building computer models of very complex multicomponent systems, which certainly include living biological organisms, requires special methods of verification and validation. Special requirements should be placed on the models that underlie decision support systems in medicine. The developers and users of these models, decision makers who use information derived from the results of these models, and those who are influenced by decisions based on such models are rightly concerned about whether the model and its results are "correct". The key problem is the high dimension of the parameter vector that determines the properties of a biological tissue or organ, when the properties of a biological tissue are radically heterogeneous. Within the medical norm, the parameter spread is very large and a time factor may appear in addition to the dependence on spatial coordinates. The determination of these physico-chemical characteristics for a living organism is difficult due to technical reasons, and is often impossible due to legal restrictions. We cannot use standard physical methods to perform a validity check for a computational model based on a physical object. The natural parameter spread of the biosystem requires diligent simulations and the construction of distributions that are compatible with medical measurements.

Based on a sequential comparison of classification results for two datasets we are developing an iterative method of verification and validation of simulation models. The first dataset contains data of medical measurements of brightness temperatures at 18 different points on the surface of mammary gland and inside the biological tissue. The second dataset contains the results of numerical modeling of the brightness temperature distribution based on a complex mathematical model of the heat dynamics and the electromagnetic field distribution inside the biological tissue. We rely on the so-called Objective Approach \cite{Sargent2014}, making a comparison using statistical tests and procedures. As a result, we have the ability to form a set of numerical simulations, which is statistically close to the medical measurement dataset. We have demonstrated a sufficiently rapid convergence of the iterative approach.

%
%
%
%

\end{document}